# Systematic Review for AI-based Language Learning Tools

**Jin Ha Woo[1], Heeyoul Choi[2]**

[1]*Arizona State University, Tempe, AZ*
[2]*Handong Global University, Pohang, South Korea*

## Abstract

The Second Language Acquisition field has been significantly impacted by a greater emphasis on individualized learning and rapid developments in artificial intelligence (AI). Although increasingly adaptive language learning tools are being developed with the application of AI to the Computer Assisted Language Learning field, there have been concerns regarding insufficient information and teacher preparation. To effectively utilize these tools, teachers need an in-depth overview on recently developed AI-based language learning tools. Therefore, this review synthesized information on AI tools that were developed between 2017 and 2020. A majority of these tools utilized machine learning and natural language processing, and were used to identify errors, provide feedback, and assess language abilities. After using these tools, learners demonstrated gains in their language abilities and knowledge. This review concludes by presenting pedagogical implications and emerging themes in the future research of AI-based language learning tools.

# Ⅰ. Introduction

The complex process of language learning involves a variety of factors leading to numerous possible outcomes. In particular, language learning is influenced by the degree of acculturation, amount of comprehensible input, attentiveness to L2 features/aspects, and opportunities for meaningful negotiation and production [1]. As people learn additional languages, they make intentional efforts to develop their language knowledge and abilities. This process involves understanding language components (e.g., grammar, vocabulary) that will help them to master the four core skills (reading, writing, listening, speaking) and related skill aspects (e.g., pronunciation) [2]. Over the past two decades, the Complex Dynamic Systems theory and the multilingual turn have led to a greater emphasis on individual learner differences in language learning [3]. In response to an emphasis on individualized learning, Intelligent Computer Assisted Language Learning (ICALL) clearly emerged as a subfield of Computer Assisted Language Learning (CALL) [4]. The ICALL subfield focuses on applying AI concepts, techniques, algorithms, and technologies to CALL, especially natural language processing (NLP), user modeling, expert systems, and intelligent tutoring systems.

Based on existing reviews related to AI in language learning, there has been a focus on developing tutoring systems, writing assistants, virtual reality environments, chatbots, and other types of adaptive learning systems/software [5]. The main intent of these tools has been to generate personalized and customizable learning experiences for the purposes of optimizing language learning by increasing autonomy, motivation, engagement, and effectiveness [5]-[6]. For instance, NLP-based tutoring systems are designed to provide tailored feedback, recommendations, and materials [4]. Recently, with the rapid development of AI, these tools can meticulously adapt content in real-time to the learning pace, preferences, and needs (e.g., cognitive, affective, social) of each user [4]-[5].

Despite the immense potential of AI in language learning, there have been concerns regarding insufficient privacy, information, and teacher preparation. Foremost, as data collection is essential to AI development, there is a need to reinforce privacy policies and informed consent practices [4]. Also, to address the lack of evidence verifying the language learning effectiveness of AI, efforts should be made to acquire information on the pedagogical effects and learner perceptions of AI-based language learning tools [5]. With this information, teachers can gain a deeper awareness of available AI-based tools which will enable them to facilitate the use of these tools effectively and appropriately.

Presently, there is a lack of comprehensive reviews on available AI-based language learning tools and the pedagogical effects and learner perceptions of these tools. Existing reviews related to AI in language learning have focused on a specific type of AI-based tool or the overall impact of AI on the future of language education [4]-[5]. Thus, an in-depth overview is needed on recently developed AI-based language learning tools, including information on the types of utilized AI technology like NLP or machine learning (ML) and the targeted language skill areas (e.g., reading, speaking) [7].

In the next section, we describe our methodological approach, the research questions and the systematic review guidelines. Then, we present our findings based on our analysis of the relevant literature. Finally, we conclude by discussing plans for future research.

# Ⅱ. Methods

## 2-1 Research Aim and Study Design

This review seeks to identify trends in the development of AI-based language learning tools and provide detailed information on these tools. To address the concerns regarding insufficient information and teacher preparation identified in the existing reviews, we aimed to answer the following questions:: What types of AI tools have been developed for various target language skill areas? How have these tools impacted language learning?

To answer these questions, we analyzed the contents of articles published in peer-reviewed journals from 2017 to 2020. We decided to narrowly focus on this three-year period because of the recent acceleration in mainstreaming AI into education. In 2017, the global venture capital investment in AI reached $1047 billion, and there has been a massive growth in AI development leading to the implementation of innovative techniques, algorithms,

approaches, and models [8].

The review was conducted based on the Preferred Reporting Items for Systematic Reviews and Meta-Analysis (PRISMA) guidelines, designed to transparently report on the purposes, methods, and findings of studies [9]. Following the PRISMA guidelines, we went through the process of identifying and screening studies on AI-based language learning tools, as shown in Figure 1.

Foremost, we identified studies using the Education Resources Information Center (ERIC), Scopus, and Web of Science database. These reputable databases were selected because of their accessibility and wide coverage of academic journals related to education. To identify trends in the development of AI-based language learning tools, we conducted an initial search using the following terms: "(artificial intelligence) AND ((language learning) OR (second language learning) OR (foreign language learning) OR (EFL) OR (ESL))". The search criteria were limited to peer-reviewed academic journals in English published from 2017 to 2020. As this initial search identified only 39 results, we conducted a follow-up search using the following specific terms related to AI and language learning: "(ICALL) OR ((language learning) AND ((natural language processing) OR (machine learning) OR (intelligent tutor) OR (bot) OR (virtual assistant))) OR ((artificial intelligence) AND ((language education) OR (language development)))". The search criteria were limited to peer-reviewed academic journals in English published from 2017 to 2020. The follow-up search identified 415 results.

For the screening process, we began by removing results that were duplicated or could not be accessed through our institution system. Then, we examined the relevancy of the title and abstract and removed results that did not mention utilizing AI for language learning. After fully reviewing the remaining 62 results, we removed six studies that were non-empirical and three studies with insufficient information on the methods and findings. Finally, 53 articles were included in this mixed-methods systematic review.

### 2-2 Data Analysis

For the final sample of 53 articles, we developed a spreadsheet template with categories quantitatively or qualitatively describing various aspects of each article. To better understand the characteristics of the articles, we compiled quantitative descriptions of the publication year, study design (methodology, location, target language/language skill), and participants (grade, target language proficiency level). The target language skills covered in this review include language components (grammar, vocabulary), macro-skills (listening, speaking, reading, writing), and a related skill aspect (pronunciation). In addition, to better understand the types of tools that have been developed and the impact of these tools on language learning, we compiled qualitative descriptions of the utilized AI technology (e.g., NLP, ML) and findings (pedagogical effects and learner perceptions).

To maintain consistency in the qualitative descriptions, we reviewed 10 articles at a time and took independent notes on the technology and findings. Then, we met to discuss and compile our notes. Through these discussions, we refined the spreadsheet categories, adjusted the level of detail in our notes, and worked towards identifying possible trends in the data.

In examining the distribution of studies across publication years (2017-2020), we found that many of the studies (N = 35, 66%) were published in 2019 or 2020. Moreover, in examining the distribution of the study design, we noticed that a significant number of studies were quantitative or mixed (N = 50, 94%), targeted the English language (N = 37, 69%), and targeted the speaking and listening (N = 14, 26%), writing (N = 11, 21%), or pronunciation (N = 11, 21%) levels. A detailed breakdown of the study design (location, target language/skill) and participants (grade, target language proficiency level) can be found in Figure 2 to Figure 6.

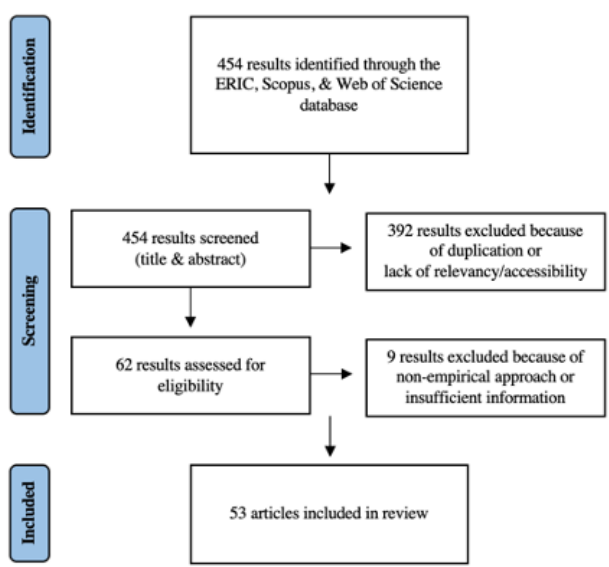

**Fig. 1**. PRISMA chart about the review phases [9].

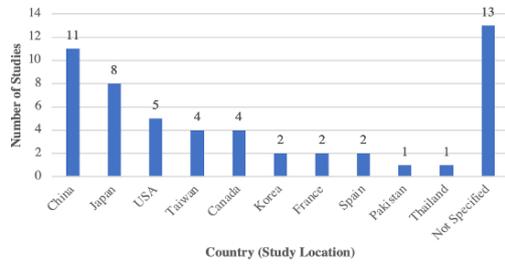

**Fig. 2.** Studies that took place in countries

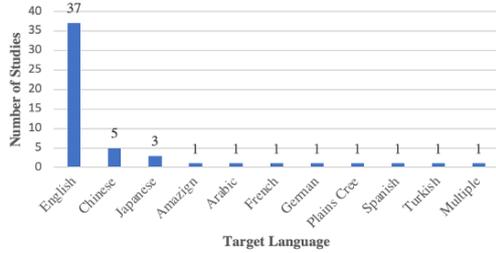

**Fig. 3.** Studies targeting various languages

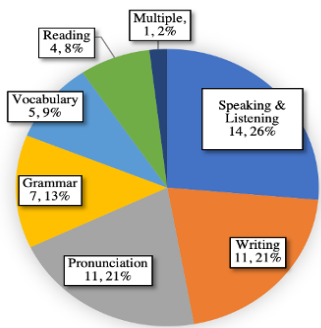

**Fig. 4.** Targeted language skill area

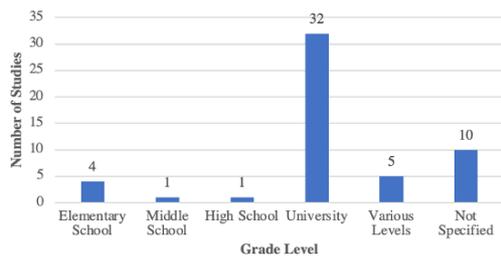

**Fig. 5.** Participants' grade level

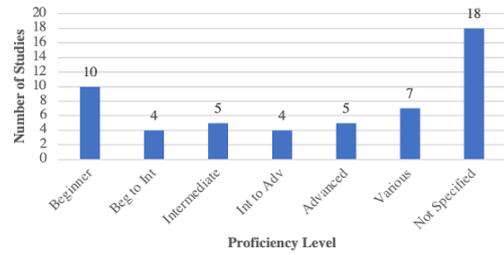

**Fig. 6.** Participants' target language proficiency

## Ⅲ. Results

### 3-1 Types of AI Tools Developed for Language Skill Areas

In response to the question, 'what types of AI tools have been developed for various target language skill areas?', we provide an overview of the tools that have been developed for each of the aforementioned skill areas (speaking, listening, writing, pronunciation, grammar, vocabulary, and reading) with the type of tool (e.g., robots, mobile applications, and vitual assistants) and AI technology.

#### 1) Speaking and Listening

In the final sample, 14 studies (26%) identified a tool targeting the speaking and listening skill. Four studies explored the potential of using intelligent personal assistants like Alexa by examining comprehensibility, usability, and improvements in listening comprehension, speaking proficiency, and willingness to communicate [10]-[13]. Additionally, programmable robots were used in group conversations ]14], and a neural network (NN)-based dialogue system was used for free conversation practice [15]. An NN-based multimodal dialog system was also developed to holistically assess spoken language in terms of delivery, content, vocabulary, and grammar [16].

#### 2) Writing

Eleven studies (21%) identified a tool targeting the writing skill area. The tools included machine translators [17], software for free-form writing [18], and a blended course with automated feedback on area. The results of the descriptive statistics also revealed that a substantial number of the studies focused on university students (N = 32, 60%). Although more studies focused on beginners (N = 10, 19%), there was a relatively even distribution across proficiency self-correcting tasks [19]. There were also specialized systems focused on citations and referencing [20] and classifying sentences into rhetoric categories [21]. These tools incorporated the Latent Semantic Analysis, random forests, support vector machines (SVM), and Naïve Bayes classifiers in ML.

### 3) Pronunciation

Eleven studies (21%) identified a tool targeting the pronunciation skill area. In 7 of these studies, deep learning algorithms [22] were used. Pronunciation diagnosis, training, and evaluation systems were developed using the attention mechanism and various types of NN (e.g., convolutional, long-short term memory) [23]-[27]. For instance, a multimodal system illustrating speech features [28] and an interactive tool generating personalized voice models [29] have recently been developed.

### 4) Grammar

Seven studies (13%) identified a tool targeting the grammar area. These tools included games, applications, immersive environments, and intelligent systems that utilized NN, ML, and NLP. For example, to create customized study plans, NN modeling was used to predict grammatical challenges that learners may encounter based on their first language [30]. For the applications and systems, word segmentation, syntactic parsing, and the finite state transducer in NLP were used to generate feedback [31]-[32].

### 5) Vocabulary

Five studies (9%) identified a tool targeting the vocabulary area. These tools included systems, platforms, robots, games, and mobile applications that have been developed using ML (e.g., conditional random field models) and NLP. For instance, in an ICALL platform, part-of-speech (POS) annotation and syntactic parsing in NLP were used to visually enhance targeted vocabulary items by automatically generating multiple-choice gaps [33]. One study (2%) identified a mobile app with adaptive learning technology that targeted both the vocabulary and grammar area [34].

### 6) Reading

The remaining four studies (8%) identified intelligent systems targeting the reading skill area. These systems utilized ML to diagnose reading problems and push appropriate resources. Additionally, an ML model was developed to identify pedagogical factors distinguishing high-achieving from low-achieving readers to improve ESL reading instruction [35].

In summary, the AI tools identified in the 53 studies targeted diverse language skill areas and incorporated many types of features, configurations, and capabilities. For many of these tools, ML and NLP were incorporated into the configuration. While NLP techniques (e.g., POS annotation, language modeling and machine translation [36]) were used more often for the writing, grammar, vocabulary, and reading tools, NNs were used more often for the speaking, listening, and pronunciation tools.

### 3-2 Impact of AI Tools on Language Learning

In response to the question, 'how have these tools impacted language learning?', we provide an overview of the pedagogical effects and learner perceptions of the AI-based tools identified in the previous section. For each skill area, we summarize the intended purposes of these tools as well as the changes in the language learning processes (e.g., experiences, development of abilities/attitudes/ knowledge) demonstrated by the learners and perceptions self-reported by the learners after using these tools.

To enhance language learning processes, AI tools were used to identify errors, provide feedback, push resources, and assess/evaluate language abilities. After using these tools, the learners demonstrated gains in their language abilities, attitudes, knowledge, and use. In general, the learners perceived these tools as effective, efficient, accurate, easy to use, and useful/helpful for language learning. Overall, the learners reported having interesting, enjoyable, and satisfactory experiences with these tools.

### 1) Correcting Errors in Grammar

AI-based grammar tools identified errors and provided on-topic feedback [31]-[32]. By using these tools, the learners were able to use English articles more accurately [37] and experience a greater sense of immersion, presence, and realism while learning [38]. Out of the 45 participants in [24], the 15 participants who used a digital game demonstrated significant gains on writing tasks eliciting the use of English articles (p = 0.000, Pre-test M = 61.6, Post-test M = 73.866, Delayed Post-test M = 84.800). The digital game-based group significantly outperformed (p = 0.000) the participants in the cloze exercise group (N = 15) and cloze exercise group with corrections (N = 15) on the writing tasks. In regard to perceptions, the learners viewed these tools as effective, efficient, accurate, enjoyable, satisfactory, and easy to use. The learners also noted that the tools adequately represented their course materials and helped them achieve their language learning outcomes [32], [39].

2) Assessing and Evaluating Conversations

For speaking and listening, AI tools were used to assess speaking abilities, evaluate conversations, and provide appropriate responses in open conversations [15], [40]-[41]. With these tools, the learners became more confident, willing, and less anxious about speaking in English [13], [42]. The learners also demonstrated gains in listening and speaking in terms of pragmatics, cohesion, word concreteness, and use of grammatical patterns [11], [14], [43]-[45]. Regarding perceptions, the learners indicated that the tools were easy to use, authentic, comprehensible, and useful for language learning [10]-[12], [41], [43], [45]. The 29 participants in [10] experimented with Google Assistant (GA) for an hour and then completed a 5-point Likert-scale questionnaire investigating the potential of using GA for language learning. Based on the items with a mean score above a four, the participants felt that GA could boost motivation to improve English listening abilities (M = 4.24) and speaking fluency (M = 4.00), reduce stress when practicing English listening (M = 4.17) and speaking (M = 4.07), improve English listening comprehension abilities (M = 4.28), and become an enjoyable hands-on tool to use (M = 4.07).

3) Suggesting Words

For vocabulary, AI tools automatically detected Japanese expressions for the purpose of providing morphological analyses and example sentences [46]. After using these tools, the learners demonstrated gains in emotion word use and semantic knowledge of phrasal verbs [33], [47]. After a 3-week treatment with an ML-based emotion synonym suggestion system, the 33 participants in [47] demonstrated significant gains (p < 0.01) on writing tasks evaluating emotion word use with an average score increase of 1.77 (avg. pre-test score = 3.41, highest possible score = 6). In regard to perceptions, the learners generally viewed these tools as interesting, easy to use, useful, and helpful for language learning [48]-[49]. Furthermore, in a survey administered by [50], the learners indicated satisfaction with various aspects of a serious language game (e.g., accessibility, skills acquisition, game mechanics, challenge/reward balance).

4) Improving Writing

AI-based writing tools identified errors, provided feedback, assessed writing abilities, and facilitated process-based academic writing [18], [20], [51]-[53]. With these tools, the learners were able to reduce plagiarism, increase editing/revising time, and correct rhetorical function, lexical, and grammatical errors [17], [20]-[21], [54]. Following the use of a machine translator for a writing task involving multiple drafts, the 34 participants in [17] demonstrated significant gains in their writing score (p = 0.000, initial draft M = 3.76, final draft M = 4.56) and decreases in the number of lexical (initial M = 5.97, final M = 3.82) and grammatical errors (initial M = 15.67, final M = 9.82) (p = 0.000). After using a feedback system, the learners also demonstrated significant improvements in their essay drafts in terms of the organization, structure, coherence, supporting ideas, and conclusion [54]. Furthermore, regarding perceptions, the learners stated that these tools were effective, easy to use, and useful/helpful for language learning [17], [20]-[21], [51], [55]. The learners also stated that these tools helped them to identify their writing strengths/weaknesses and increase their writing knowledge through detailed comments [51].

5) Improving Fluency in Pronunciation

For pronunciation, tools were used to detect mispronunciations and recognize speech for diagnosis, assessment, and evaluation [23], [25]-[27], [56]. These tools helped the learners improve their fluency, comprehensibility, tone, and pronunciation accuracy [24], [28]-[29], [57]. With regard to perceptions, the learners described these tools as interesting, easy to use, and helpful for fluency, intonation, and tone training [24], [29], [57].

6) Personalized Tutoring

AI-based reading tools classified learners, assessed reading abilities, and pushed resources [35], [58]-[59]. For example, an adaptive learning system pushed resources that corresponded to learner characteristics (e.g., reading abilities, cognitive styles, learning objectives) [59]. After using an intelligent tutoring system, the learners demonstrated improvements in essential academic reading skills, including main idea, text structure, and inference [58].

To summarize, the AI tools identified in the 53 studies have had a positive impact on language learning and have been well-received by learners. These tools identify errors, provide feedback, and assess language abilities. Moreover, they have helped learners to build on their language abilities and enhance their learning experiences. In particular, for grammar and speaking/listening, the AI tools positively impacted psychological factors such as confidence, anxiety, immersion, and presence. By placing a greater emphasis on psychological factors, teachers can help learners to increase their comprehensible input and attentiveness to L2 features/aspects leading to greater gains [60].

## Ⅳ. Discussions

Based on the reviewed studies, it is clear that various AI tools targeting the speaking, listening, writing, pronunciation, grammar, vocabulary, and reading area have been developed. After using these tools, learners have demonstrated improvements in their language skills/knowledge and perceived these tools to be useful for language learning.

Since learners work directly with these AI tools, learners and teachers need to become familiarized with the fundamentals of commonly utilized AI technologies. By bringing awareness to a wide array of available AI tools, teachers can guide learners to select the most appropriate tools for their language learning preferences and needs. For example, learners who prioritize familiarity and accessibility can work with existing technologies such as Alexa and Google Translate. On the other hand, learners who prioritize collaborative learning can work with innovative tools such as NAO, a programmable robot for group conversations.

Additionally, since there are tools targeting very specific language abilities, teachers who are unfamiliar with AI can begin by experimenting for an activity without a long-term commitment. As teachers experiment with new technologies, they will gain the knowledge and experiences needed to innovatively implement and facilitate student-centered technology use.

Lastly, learners have generally perceived many of these tools as effective, interesting, easy to use, and helpful. However, to determine the types of AI tools that work best for specific types of learners, teachers and researchers can administer open-ended surveys and conduct interviews to better understand the reasoning behind their perceptions.

This review was limited to articles in peer-reviewed academic journals published in English from 2017 to 2020. Moreover, only three databases were used to identify articles, and some relevant articles could not be accessed through our institution system. This could have been the reason why the studies were skewed towards university students learning English in China or Japan.

Researchers plan to make systems more effective and improve the accuracy of detection/recognition capabilities. Additionally, researchers need to incorporate pedagogical knowledge by collaborating with teachers and examining in-class use of these tools. Researchers have also emphasized the need to conduct studies that include diverse learners, examine the long-term impact on learners, and verify contributions to language learning. Lastly, there are plans to expand their user bases by making the AI tools available in languages other than English and adding more materials (e.g., feedback, exercises) for a wider range of proficiency levels. Also, there is a need to monitor how recently developed AI tools including large-scale deep learning models including BERT and GPT-3 will be applied to language learning in the near future.